\documentclass[runningheads]{llncs}
\usepackage[T1]{fontenc}

\usepackage{graphicx}
\usepackage[round]{natbib}
\usepackage{bm}
\usepackage{url}
\usepackage{xspace}
\usepackage{amsmath,amsfonts,amssymb,amscd,amsthm,xspace,mathtools}
\usepackage[hidelinks]{hyperref}
\usepackage{xcolor}
\usepackage[ruled,vlined]{algorithm2e}

\DeclareMathOperator*{\argmax}{arg\,max}

\usepackage[colorinlistoftodos]{todonotes}
\definecolor{citrine}{rgb}{0.89, 0.82, 0.04}
\definecolor{blued}{RGB}{70,197,221}
\definecolor{applegreen}{rgb}{0.55, 0.71, 0.0}
\definecolor{flame}{rgb}{0.89, 0.35, 0.13}

\begin{document}

\title{State and Action Factorization in Power Grids}

\author{
Gianvito Losapio\inst{1,*}
\and
Davide Beretta\inst{1}
\and
Marco Mussi\inst{1}
\and
Alberto Maria Metelli\inst{1} 
\and
Marcello Restelli\inst{1}
}

\authorrunning{Losapio et al.}

\institute{\textsuperscript{1}Politecnico di Milano, Milan, Italy \\
\email{\inst{*}gianvito.losapio@polimi.it}}

\maketitle            

\begin{abstract}
The increase of renewable energy generation towards the zero-emission target is making the problem of controlling power grids more and more challenging. The recent series of competitions Learning To Run a Power Network (L2RPN) have encouraged the use of Reinforcement Learning (RL) for the assistance of human dispatchers in operating power grids. All the solutions proposed so far severely restrict the action space and are based on a single agent acting on the entire grid or multiple independent agents acting at the substations level. In this work, we propose a domain-agnostic algorithm that estimates correlations between state and action components entirely based on data. Highly correlated state-action pairs are grouped together to create simpler, possibly independent subproblems that can lead to distinct learning processes with less computational and data requirements. The algorithm is validated on a power grid benchmark obtained with the Grid2Op simulator that has been used throughout the aforementioned competitions, showing that our algorithm is in line with domain-expert analysis. Based on these results, we lay a theoretically-grounded foundation for using distributed reinforcement learning in order to improve the existing solutions.
\keywords{Power Grids \and Factorization \and Reinforcement Learning}
\end{abstract}

\section{Introduction}
\label{sec:introduction}
Distributing electricity from generation sources to end-users is an extremely complex task. Transmission System Operators (TSOs) across the world must ensure a safe supply of electricity through power grids, meeting the demand at all times while preventing blackouts. Power grids are controlled from control centers by \textit{human dispatchers}, with remote observability to all transmission network elements. Their role is to monitor the electricity network 24 hours per day, 365 days per year. They must constantly keep the network within its thermal limits, frequency ranges, and voltage ranges by taking \textit{remedial actions} on network elements such as lines and substations via remote control command \citep{kelly2020reinforcement}.
Nowadays, the increase of renewable energy generation towards the zero-emission target is making the problem of controlling power grids more and more challenging. The energy production by renewable generators is constantly fluctuating due to weather conditions, making power grid operations a stochastic control problem. Existing software, computational methods, and optimal power flow solvers
are not adequate for real-time network operations on short temporal horizons in a reasonable computational time \citep{kelly2020reinforcement, serre2022reinforcement}. In order to overcome these complexities, TSOs need new methods for controlling power grids. Among emerging techniques, Reinforcement Learning (RL) has been explored in several works over the last years with promising results \citep{dorfer2022power, chauhan2023powrl, van2023multi, lehna2024hugo}.

RL is the special class of Machine Learning (ML) methods dealing with sequential decision-making problems \citep{sutton2018reinforcement}. Controlling a power grid with RL means learning an optimal control policy from data. The recent success of RL in games \citep{silver2017mastering}, robotics \citep{haarnoja2024learning}, and industrial processes \citep{luo2022controlling} - just to name a few - is due to recent algorithms powered by deep learning that can efficiently handle complex and heterogeneous data. The main advantage of using RL is that it can identify and capitalize on under-utilized, cost-effective actions that human dispatchers and traditional solution techniques are unaware of or unaccustomed to, resulting in more efficient control of the power grid by learning effective relations on stochastic data. In problems such as power grid control, there is a huge quantity of observed variables about the grid (e.g., power production, load consumption, power flow over transmission lines) and an impressive amount of control actions that human dispatchers can take (e.g., topological changes, generator dispatching). With this large state and action space, RL algorithms suffer a problem known as the \textit{curse of dimensionality}, i.e., the amount of data/computation required to achieve a good solution may be out of reach.

Distributed Reinforcement Learning (DRL) algorithms can be considered to mitigate this problem by distributing the learning process among multiple agents~\citep{zhang2021multi}. In this framework, each agent can observe just a limited part of the state space and take only a small number of actions, but all the agents cooperate to achieve a common goal. The main idea of DRL is thus breaking the complexity of the original RL problem by creating smaller and simpler subproblems. Designing subproblems is, therefore, a crucial task that may critically affect the performance of the learning algorithms. Obtaining a factorization of the state and the action space of a power grid is an extremely challenging task since localized actions may have side effects on a distant portion of the grid, thus requiring including in the observation/action space of an agent information about other elements of the grid despite their spatial distance \citep{marot2018guided}.

\paragraph{Original Contribution.} In this work, we propose an algorithm for state and action factorization of the power grid control problem that can be used to exploit the benefits of DRL methods compared to traditional RL. The algorithm is \textit{domain-agnostic}, meaning that it can be applied to any complex decision-making problem beyond power grids as it does not require any \textit{a priori} knowledge of the domain of the problem. Correlations between state and action components are entirely computed on data by using mutual information, an information-theoretic measure that is used to assess how much a state or an action variable (input variables) can be used to predict the evolution of future values of the state variables (target variables). Highly correlated state-action pairs are grouped together to create simpler, possibly independent subproblems that can lead to distinct learning processes, e.g., over different portions of the grid with fewer data requirements.
The algorithm is validated on a power grid benchmark obtained simulating realistic production/consumption profiles with the open-source simulator Grid2Op\footnote{\url{https://github.com/rte-france/Grid2Op}}, showing that our algorithm, despite being domain-agnostic, is in line with domain-expert analysis. Based on these results, we propose to use then DRL algorithms to solve subproblems identified by the computed factorization with the objective of improving the existing solutions for power grid control based on traditional RL.

\paragraph{Paper Structure.} The paper is organized as follows: Section~\ref{sec:relatedworks} presents related works, Section~\ref{sec:problemformulation} contains a general formulation of the problem of state and action factorization, Section~\ref{sec:algorithm} contains our algorithm with the associated pseudocode, Section~\ref{sec:experiments} shows the experiments, and finally Section~\ref{sec:conclusions} concludes the paper with possible future work.

\section{Related Works}
\label{sec:relatedworks}

\paragraph{RL for power grids.} The recent series of competitions Learning To Run a Power Network \citep[L2RPN,][]{kelly2020reinforcement, serre2022reinforcement} have encouraged the use of RL for the assistance of human dispatchers in operating power grids. Originally developed by Reseau de Transport d’Electricitée (RTE), the French TSO, its aim has been to promote investigation into the network operation problem in a competitive context. For that purpose, RTE developed the open-source Grid2Op simulator to model and study a large class of power system-related problems and facilitate the development and evaluation of agents that act on power grids. The Grid2Op simulator is a flexible tool, allowing researchers to accurately simulate power system dynamics for different networks while interacting with the environment through different types of actions.

Most of the RL methods developed to control power grids are thus related to the L2RPN competitions. An overview of the solutions proposed during the first edition is reported in \citep{marot2021learning}. The winning solutions of the last two editions are presented in \citep{dorfer2022power} and \citep{artelys2023}, respectively. All the solutions proposed so far severely restrict the action space and are based on a single agent acting on the entire grid. For instance, the authors of \citep{dorfer2022power} and \citep{artelys2023} reduce the action space to a small set of topological actions (i.e., changing bus connections at a few substations) and rely on standard optimization solvers for other types of actions such as generator redispatching. RL is thus limited in its capacity, preventing the original intention of exploring the entire action space.

Subsequent works have further explored RL algorithms beyond the scope of the competitions yet with the same limitations described above \citep{chauhan2023powrl, lehna2024hugo}. On the other hand, \citep{van2023multi} proposed for the first time a multi-agent RL approach as a solution to reduce the action space by creating an agent for each substation and specializing it on its own topological actions only. A rule-based method based on the line overloads decides which agent has to act. Similarly, \citep{manczak2023hierarchical} presents a hierarchy of agents with different combinations of high-level and low-level agents. For instance, one possibility is having a low-level RL agent that uses action masking to select a subset of actions dictated by a higher-level agent, which, in turn, is based on heuristics or RL algorithms. There are two main limitations of these approaches: ($i$) the observation space is not reduced as each agent observes the entire grid, ($ii$) the number of low-level agents is related to the number of configurable substations, thus requiring a large number of agents on grids of increasing size.

With our method, we would like to overcome such limitations by distributing the learning process across a smaller number of agents, each taking care of a specific subproblem that has been identified in the data-driven decomposition of the original problem (for instance, controlling a zone of the power grid with several substations). Each agent has thus a reduced action space and, at the same time, can only observe the relevant state variables of its own subproblem, resulting in a simpler learning problem.

\paragraph{Power grid segmentation.} The segmentation of large-scale power grids into zones is crucial for human operators when controlling the grid in real-time. Power grids are usually segmented into static zones that are redefined every year to study the grid efficiently in real-time. Typically, the segmentation has been computed using analytical methods that have been more extensively explored in the field of power systems \citep{marot2018expert}. Only recently, a data-driven approach using machine learning has been proposed \citep{marot2018guided}. By simulating the effects of a specific intervention of human operators (i.e., line disconnections on the other lines), the authors were able to create an adjacency matrix of a directed graph of lines on which they executed a graph clustering algorithm. The resulting clusters were used to segment the power grid, showing interesting results in different realistic scenarios. 

Our method has in principle a similar objective of power grid segmentation but it is designed to be more general, i.e., to segment the entire state and the action spaces with respect to any kind of intervention and independently from the physical configuration of the grid. Most notably, our method is meant to be domain-agnostic and widely applicable to any complex decision problem.

\section{Problem Formulation}
\label{sec:problemformulation}

A power grid is represented by an undirected graph with nodes being substations and edges being transmission lines. Generators and loads are directly connected to substations. Energy demand requested by loads must be satisfied by the production of generators at any time, but at the same time, each transmission line can only carry a limited amount of energy. Human dispatchers constantly monitor the grid and remotely operate \textit{remedial actions} that ensure safe working conditions of the grid, such as topology changes at the substations and dispatching of generators.

As a consequence, from the perspective of a dispatcher, power grid operations can be modeled as a sequential decision-making problem. At each time step, the grid is represented by a set of variables that we may call state (e.g., information about generators, lines), and actions are taken causing the grid to transition to a new state and producing an operational cost to the TSO related to different factors (e.g., energy lost due to the Joule effect, cost of re-dispatched energy).

Markov Decision Processes (MDPs, \citealt{puterman1990markov}) offer a well-studied mathematical framework to model sequential decision-making problems and are used for the formulation of RL problems. An MDP is formalized as a tuple $\mathcal{M} \coloneqq \left\langle \mathcal{S}, \mathcal{A}, P, R, H\right\rangle$, where $\mathcal{S}$ is the set of states the environment can assume, $\mathcal{A}$ is the set of actions the agent can execute, $P: \mathcal{S} \times \mathcal{A} \times \mathcal{S} \rightarrow [0,1]$ is the stochastic transition function ($P(s'|s,a)$ being the probability of moving to next state $s'$ when performing action $a$ in state $s$), $R: \mathcal{S} \times \mathcal{A} \rightarrow \mathbb{R}$ is the reward function ($R(s,a)$ being the reward the agent gets when performing action $a$ in state $s$), and $H \in \mathbb{N}$ is the planning horizon.

In MDPs, both the transition probability and the reward function depend only on the current state and action, as the state and action histories are irrelevant (\emph{Markov property}). 
The agent chooses actions according to a \emph{policy} $\pi: \mathcal{S} \times \mathcal{A} \rightarrow [0,1]$ that maps states to stochastic actions, $\pi(a|s)$ being the probability that the agent chooses action $a$ when the environment is in state $s$. Finding an optimal solution to an MDP means searching for a policy $\pi^*$ that maximizes the expected sum of the rewards obtained by the agent during the $H$ steps of interaction with the environment, formally $\pi^* \in \argmax_{\pi} \mathbb{E}_{\pi} \left[ \sum_{t=1}^{H} R(s_t, a_t) \right]$. Rewards are here considered equally important over time (\textit{undiscounted finite-horizon} MDP).

\paragraph{Factorizable Structure.} In this work, we consider a \emph{factored} structure for the MDP $\mathcal{M}$ in which the state and action vectors have components
\begin{align*}
&\mathbf{s} = (s_{1}, s_{2}, \dots, s_n) \in \mathcal{S}, \\
&\mathbf{a} = (a_{1}, a_{2}, \dots, a_m) \in \mathcal{A}.
\end{align*}
Moreover, we suppose that our MDP $\mathcal{M}$ can be seen as composed of $K$ independent MDPs $\mathcal{M} = (\mathcal{M}_k)_{k=1}^{K}$, with each one being defined as:

$$
\mathcal{M}_k = \Big( \mathcal{S}_k,\mathcal{A}_k, P_k, R_k, H \Big)\,.
$$

In this formulation, each state space $\mathcal{S}_k$ and action space $\mathcal{A}_k$ of the MDP $\mathcal{M}_k$ are taken as a subset of the space $\mathcal{S}$ or $\mathcal{A}$ by combining the domains of only some components of the vector $\mathbf{s}$ or $\mathbf{a}$. \footnote{In the case for example $\mathcal{M} = (\mathcal{M}_1, \mathcal{M}_2)$, we can suppose that the state of $\mathcal{M}_1$ is composed by the first two components $(s_1, s_2)$ of the state vector $\mathbf{s}$ and the action of $\mathcal{M}_1$ is composed by the first component $(a_1)$ of the action vector $\mathbf{a}$. The state space and the action space of $\mathcal{M}_1$, namely $\mathcal{S}_1$ and $ \mathcal{A}_1$, are thus obtained by combining the corresponding domains of only those components.}

This formulation seems to resemble the framework of Factored MDPs~\citep{degris2006learning}. However, we assume a different factorization of the global transition probability and of the global reward function. The underlying MDPs presents transition probabilities $P_k :\mathcal{S}_k \times \mathcal{A}_k \times \mathcal{S}_k \rightarrow [0,1]$ and the reward functions $R_k: \mathcal{S}_k \times \mathcal{A}_k \rightarrow [0,1]$. The global transition probability and reward function of $\mathcal{M}$ are defined as:
\begin{align*}
&P(\mathbf{s}' | \mathbf{s}, \mathbf{a}) = \prod_{k=1}^{K} P_k(\mathbf{s}_k{'} \,|\, \mathbf{s}_k, \mathbf{a}_k)\,, \\[3mm]
& R(\mathbf{s},\mathbf{a}) = f \left( R_1 (\mathbf{s}_1 , \mathbf{a}_1) , R_2 (\mathbf{s}_2 , \mathbf{a}_2) , \ldots , R_K (\mathbf{s}_K , \mathbf{a}_K) \right),
\end{align*}

where the variables in bold $\mathbf{s}_k \in \mathcal{S}_k, \mathbf{a}_k \in \mathcal{A}_k$ refer to the state/action vectors of the MDP $\mathcal{M}_k$ (not to be confused with the scalar components of the vectors $\mathbf{s}, \mathbf{a}$ of the original MDP $\mathcal{M}$). The term $f ( \cdot )$ refers to a fixed function that combines the rewards of all the MDPs.

It is easy to show that with this formulation we can consider $-$ without loss of generality $-$ independent policies on each MDP $\mathcal{M}_k$ to optimize the global MDP $\mathcal{M}$. The main idea of this paper is therefore to find an algorithm that returns an estimate of the state and action factorization $\big( \widehat{\mathcal{S}}_k, \widehat{\mathcal{A}}_k  \big)_{k=1}^{\widehat{K}}$ that ideally matches the true factorization $\big(\mathcal{S}_k, \mathcal{A}_k  \big)_{k=1}^{K}$. Then, we can use DRL algorithms on each of the MDPs defined by the estimated factorization and optimize the original problem $\mathcal{M}$.

We can start by defining a fully connected graph $\mathcal{G}=(V,E)$ in which

\begin{itemize}
    \item[$\bullet$] $V$ is the set of nodes containing all the state and action components from $\mathbf{s}, \mathbf{a}$ and all the next state components from $\mathbf{s'}$
    \item[$\bullet$] $E$ is the set of edges representing the interactions among components
    \begin{equation}\label{edges}
    E = \{(x_i,s'_j) \,|\, x_i,s'_j\in V\; \text{and}\; c(x_i,s'_j)\geq \delta\},
    \end{equation}
    where $c(x_i,s'_j)$ is a metric that measures how much a variable $x_i$ (state or action component) is important to predict the variable $s'_j$ (next state component), with $\delta$ being a suitable threshold.
\end{itemize}

With the above definition, the presence of an edge $(x_i,s'_j)$ in the graph $\mathcal{G}$ means that the component of the next state $s'_j$ can be predicted by the component $x_i$ of the state or action vector, thus $x_i$ and $s'_j$ belong to the same MDP. In this formulation, we can discard the weak connections by means of the threshold $\delta$, and we can obtain either an undirected or a directed graph if the metric is symmetric or asymmetric, respectively. On this graph, we can then run a clustering algorithm to divide the variables into communities $\big( \widehat{\mathcal{S}}_k, \widehat{\mathcal{A}}_k  \big)_{k=1}^{\widehat{K}}$.

\paragraph{Performance Evaluation.} In order to evaluate the performance of the predicted factorization $\big( \widehat{\mathcal{S}}_k, \widehat{\mathcal{A}}_k  \big)_{k=1}^{\widehat{K}}$  compared to the real factorization $\big(\mathcal{S}_k, \mathcal{A}_k  \big)_{k=1}^{K}$, we can suppose there is a ground-truth adjacency matrix for the graph $\mathcal{G}$ denoted as $I_{\mathcal{G}}$ that we can compare to our prediction $\widehat{I}_{\mathcal{G}}$ (computed with Eq.~\ref{edges}) by means of some suitable metric $\ell$, e.g., the Frobenius norm:

\begin{equation}\label{metric}
\ell\left(I_{\mathcal{G}}, \widehat{I}_{\mathcal{G}}\right) = \left\lVert  I_{\mathcal{G}} - \widehat{I}_{\mathcal{G}} \right\rVert_{F}^2 = \sum_{i,j} \left(I_{\mathcal{G}}[i,j] - \widehat{I}_{\mathcal{G}}[i,j]\right)^2.
\end{equation}
Therefore, based on $\ell$, it may be possible to quantify an approximation error that gives the performance of the algorithm.

\section{Algorithm}
\label{sec:algorithm}
We propose an algorithm that uses an information-theoretic measure to build the adjacency matrix $\widehat{I}_{\mathcal{G}}$ that describes the correlation among state and action variables, and finds a factorization of the original MDP $\mathcal{M}$. The pseudo-code of the procedure is reported in Algorithm~\ref{algo}.

\begin{algorithm}[t!]\label{algo}
\caption{State and Actor Factorization}
\SetAlgoLined

\vspace{1mm}

\textbf{Input}: MDP $\mathcal{M}$, Explorative policy $\pi_e$, Threshold $\delta$ \\

\vspace{2mm}

\textbf{Output}: Factorization $\big( \widehat{\mathcal{S}}_k, \widehat{\mathcal{A}}_k  \big)_{k=1}^{\widehat{K}}$

\vspace{2mm}

\textbf{Algorithm:}\\[1mm]

\textbf{1.} Collect a dataset $\mathcal{D}$ of transitions from $\mathcal{M}$ with policy $\pi_e$ \\[1mm]

\textbf{2.} Compute the adjacency matrix $\widehat{I}_{\mathcal{G}}$ approximating the mutual information on $\mathcal{D}$ and using $\delta$ as threshold \\[1mm]

\textbf{3.} Transform $\widehat{I}_{\mathcal{G}}$ into a pseudo-block diagonal matrix and define the set of clusters $\big( \widehat{\mathcal{S}}_k, \widehat{\mathcal{A}}_k  \big)_{k=1}^{\widehat{K}}$ corresponding to diagonal blocks \\[1mm]

\end{algorithm}

The first step is to collect a dataset $\mathcal{D}$ from $\mathcal{M}$ with a sufficiently exploratory policy $\pi_e$ (see, e.g., \citealt{mutti2021task}):
$$\mathcal{D} = \big\{(\mathbf{s}, \mathbf{a}, \mathbf{s'})_t\big\}_{t=1}^{T}$$
where each entry is a transition that starts from state $\mathbf{s}$, plays an action $\mathbf{a} \sim \pi_e(\cdot \,|\, \mathbf{s})$ and reaches state $\mathbf{s'}$.

We define a random variable $S'$ that represents the next state vector and a random variable $X$ that is the concatenation of the state vector and the action vector:
\[
S' = (S'_{1},S'_{2},\dots, S'_{n}),
\]
\[
X=(S_{1},S_{2},\dots,S_{n}, A_{1}, A_{2},\dots, A_{m}),
\]
Each entry of the dataset $\mathcal{D}$ is a joint realization of $X$ and $S'$. We can now define the connectivity matrix $\widehat{I}_{\mathcal{G}}$ as a matrix of size $(n \times (n+m))$ having a row for each component of $S'$ and a column for each component of $X$. Each entry $\widehat{I}_{\mathcal{G}}[i,j]$ is computed based on the Mutual Information (MI) between the component $i$ of the next state vector, $S'_i$, and the component $j$ of state-action vector, $X_j$,
\[
\widehat{I}_{\mathcal{G}}[i,j] =
\begin{cases}
1 & \,\,\, \text{if} \,\,\,  \text{MI}(S'_i,X_j) \geq \delta \\
0 & \,\,\, \text{otherwise}
\end{cases}
\]
where
\[
\text{MI}(S'_i,X_j) := \mathbb{E}\Bigg[\log {\Bigg({\frac {p(s'_i,x_j)}{p(s'_i)\,p(x_j)}}\Bigg)\Bigg]}
\]
quantifies the amount of information (or, equivalently, reduction in uncertainty) that knowing either variable provides about the other, $\delta$ is a suitable threshold. The quantity $\text{MI}(S'_i,X_j)$ cannot be computed exactly as, in practice, we do not have access to those probability distributions, but it is approximated using the dataset $\mathcal{D}$. 

At this point, the binary matrix $\widehat{I}_{\mathcal{G}}$ can be transformed into a pseudo-block diagonal matrix, arranging the columns so that variables $X_j$ that have an impact on the same components $S'_i$ are close together. Based on such diagonal blocks, we can define the set of clusters $\big( \widehat{\mathcal{S}}_k, \widehat{\mathcal{A}}_k  \big)_{k=1}^{\widehat{K}}$ and run a DRL algorithm on each corresponding MDP.

\section{Experiments}
\label{sec:experiments}
We performed two different experiments to test the effectiveness of our algorithm. In the following, each experiment is presented in a separate paragraph, with details on data collection and evaluation metrics. The estimator proposed by \citep{gao2017estimating} is used to compute the mutual information\footnote{a bias correction is applied by computing the mutual information also on a modified version of $\mathcal{D}$ in which existing correlations among variables are arbitrarily broken.}. A recursive depth-first search is called alternately on rows and columns of $\widehat{I}_{\mathcal{G}}$ to obtain a pseudo-block diagonal matrix in which the input variables $X_j$ that influence the same target variables $S'_i$ are grouped together in the same block. The threshold $\delta$ is properly tuned to maximize the effectiveness of this block diagonalization.

\paragraph{Synthetic data.} The first experiment is conducted on synthetic data generated according to a specific distribution. Each state vector $\mathbf{s}$ has $n=5$ components, whereas each action vector $\mathbf{a}$ has $m=3$ components. All components are in the range $[0,1]$. The following state-action factorization is assumed:
\[
(1): \,\mathbf{s}_1 = (s_1, s_3, s_5), \,\mathbf{a}_1 = (a_1, a_2)
\]
\[
(2): \, \mathbf{s}_2 = (s_2, s_4), \,\mathbf{a}_2 = (a_3)
\]

Starting from a state $\mathbf{s}$, every interaction consists of selecting a random action $\mathbf{a}$ and then taking each component of the next state $\mathbf{s'}$ as a copy of any state or action component of the same cluster with equal probability. For instance, the first component of the next state $s'_{1}$ will be a copy of any of $\{ s_1, s_3, s_5, a_1, a_2 \}$ since the first component of $\mathbf{s}$ belongs to cluster (1).

With a dataset of $T=10^5$ samples, the original factorization is perfectly reconstructed with $\delta$ computed as the 0.5\textsuperscript{th} quantile of each column of the matrix. The corresponding approximation error of $\widehat{I}_{\mathcal{G}}$ is $0.02$, computed as the Frobenius norm (Eq.~\ref{metric}) divided by the size of $\widehat{I}_{\mathcal{G}}$ (i.e., only $1/40$ elements of $\widehat{I}_{\mathcal{G}}$ is different from the ground truth matrix).

\begin{figure}
    \centering
    \includegraphics[width=1\linewidth]{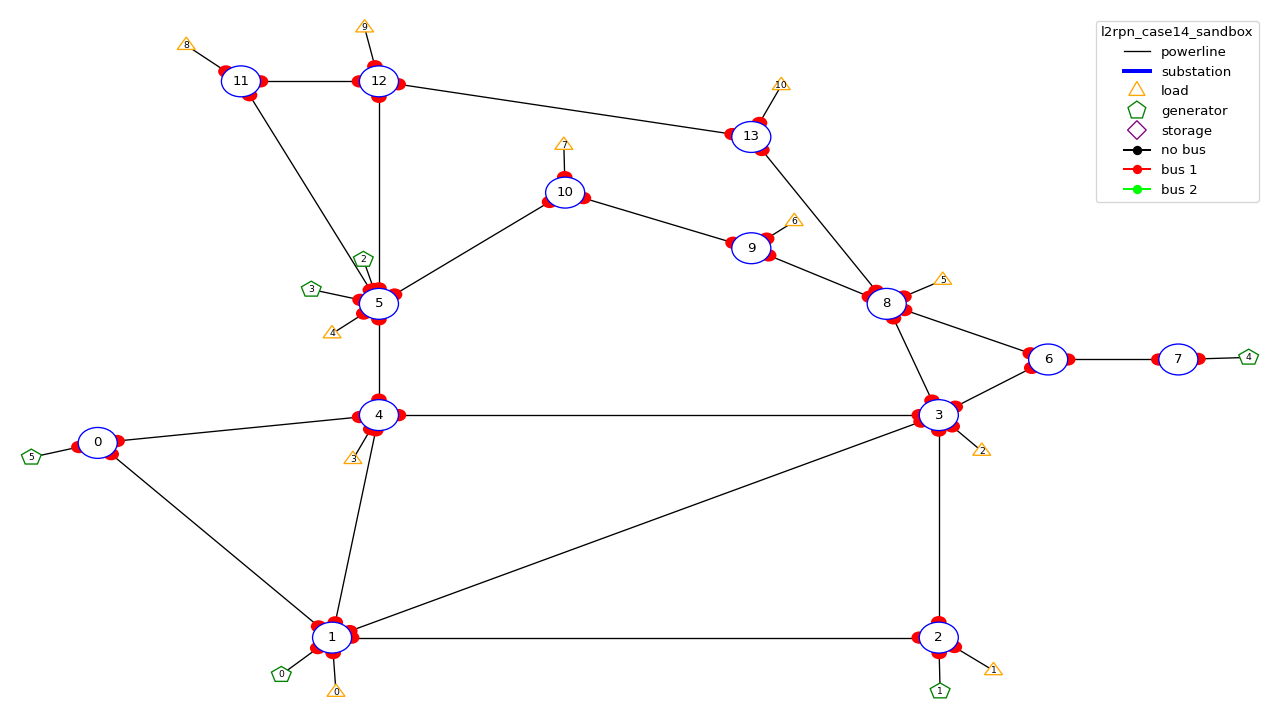}
    \caption{A layout of the grid used in the experiment with the Grid2Op simulator.}
    \label{fig:l2rpn_case14_sandbox}
\end{figure}

\paragraph{Power grid simulation.} The simulated power grid used in our experiment is shown in Figure~\ref{fig:l2rpn_case14_sandbox}. It is called \verb|l2rpn_case14_sandbox| in the Grid2Op simulator, and it is based on the IEEE case14 power grid benchmark. It counts 14 substations, 20 lines, 6 generators (of which 3 renewables), and 11 loads. 

The state space of the Grid2Op includes much information about the grid, such as generators' power production, loads power consumption, power line state (including the power flow as a fraction of capacity, denoted as $\rho$), topology configuration (bus configurations at each substation). The action space includes four types of actions: topology modification (i.e., switching bus at substations, line connection/disconnection), redispatching (i.e., requesting a change in energy production in non-renewable generators), curtailment (i.e., reducing power production in renewable generators), storage actions (i.e., storing/retrieving energy from storage).

We restrict the state space to the power flow of each power line as a fraction of its thermal capacity (variable $\rho$) and the action space to the vector of topological changes for each substation. This restriction simplifies the problem but at the same time takes into consideration two fundamental aspects of the power grid: (\textit{i}) $\rho$ is the most significant variable for the real-time monitoring of the grid because line overloading may potentially cause a cascading failure that leads to a blackout, requiring the immediate intervention of a human dispatcher; (\textit{ii}) topological actions are the only non-costly available actions and are thus preferred by TSOs.

\begin{figure}
    \centering
    \includegraphics[width=1\linewidth]{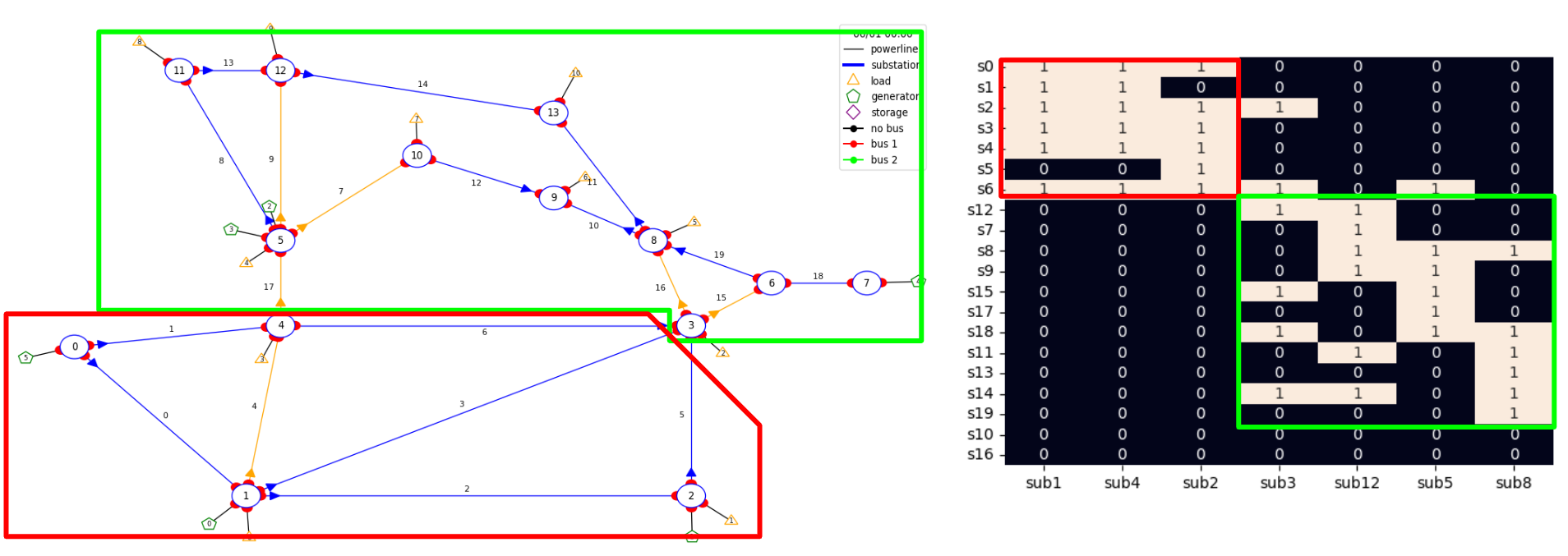}
    \caption{The factorization obtained in the Grid2Op experiment. (Right) Pseudo-block-diagonalization of the estimated portion of the adjacency matrix $\widehat{I}_{\mathcal{G}}$. (Left) Corresponding power grid segmentation.}
    \label{fig:segm}
\end{figure}

We considered only a small portion of the adjacency matrix $\widehat{I}_{\mathcal{G}}$, i.e., the columns of the action components corresponding to the substations with more than 3 connected elements\footnote{Topological actions on the remaining substations are not useful since they involve disconnections of some of their elements (which may cause disruptions on the grid).}, with the objective of having a factorization of topological actions based on their direct influence on power lines. A random policy is used to collect a dataset of $T\in[10^4,10^5]$ transitions, separately for each substation. With a threshold $\delta$ equal to the 0.7th quantile on each column, we obtained a factorization of two clusters, corresponding to a segmentation of the power grid shown in Figure~\ref{fig:segm}.

In this case, we cannot compute an approximation error since we do not have a ground truth adjacency matrix, but we can observe that our results are in line with the domain-expert analysis presented in \citep{marot2018expert, marot2018guided} that produces an analogous power grid segmentation. 

\section{Conclusion}
\label{sec:conclusions}
In this paper, we introduced an algorithm for state and actor factorization in power grids. The main advantage of our algorithm is that it is domain-agnostic in the sense that it is applicable not only to power grids but can also be extended to any complex decision-making problem as it does not require specific knowledge about the problem. In fact, the factorization is entirely based on mutual information estimated on data, which provides theoretically grounded insights on how much variables of state-action pairs are informative about each other. The results of our algorithm are in line with domain-expert analysis on a power grid benchmark obtained with an open-source simulator, demonstrating its potential applicability for power grid segmentation. Consequently, DRL algorithms can be used on the discovered subproblems to improve existing solutions for controlling power grids based on traditional optimization methods or standard RL.

Future works can include the use of correlation metrics different from mutual information, possibly reducing the number of samples required by its estimators. A theoretical analysis is also required to understand how the choice of the policy for data collection influences the data distribution and how the choice of the threshold for building the adjacency matrix can be automatized. Finally, in case the discovered factorization produces some dependencies among subproblems, it may be interesting to investigate efficient forms of communication among agents that can be applied in distributed reinforcement learning algorithms.

\vspace{0.5cm}

\noindent\textbf{Acknowledgments.} The authors acknowledge the project AI4REALNET that has received funding from European Union’s Horizon Europe Research and Innovation programme under the Grant Agreement No 101119527. Views and opinions expressed are, however, those of the authors only and do not necessarily reflect those of the European Union. Neither the European Union nor the granting authority can be held responsible for them. The authors also acknowledge the project FAIR that has been funded by the European Union – Next Generation EU within the project NRPP M4C2, Investment 1.3 DD. 341 -  15 March 2022 – FAIR – Future Artificial Intelligence Research – Spoke 4 - PE00000013 - D53C22002380006.

\bibliographystyle{unsrtnat}
\bibliography{biblio}

\end{document}